\RequirePackage{silence}
\documentclass[aps,pra,twocolumn,showpacs,amsmath,amssymb,preprintnumbers,superscriptaddress,10pt,longbibliography,showkeys]{revtex4-2}

\usepackage{graphicx}
\graphicspath{{figures/}}
\usepackage{SIunits}
\usepackage{hyphenat}
\usepackage[bookmarks=false]{hyperref}
\usepackage{color}
\usepackage[capitalise]{cleveref}
\crefname{section}{Sec.}{Secs.}
\Crefname{section}{Section}{Sections}
\usepackage{tabularray}

\definecolor{pink}{RGB}{255,0,255}

\definecolor{green}{RGB}{0,128,0}

\definecolor{turquoise}{RGB}{0,200,200}

\begin{document}

\title{Fiber-optic power limiter device based on carbon nanotubes}

\author{Ekaterina~Borisova}
\affiliation{Russian Quantum Center, Skolkovo, Moscow 121205, Russia}

\author{Anastasiya~Ponosova}
\affiliation{Russian Quantum Center, Skolkovo, Moscow 121205, Russia}
\affiliation{NTI Center for Quantum Communications, National University of Science and Technology MISiS, Moscow 119049, Russia}

\author{Natalia~Arutyunyan}
\affiliation{Prokhorov General Physics Institute of the Russian Academy of Sciences, Moscow 119991, Russia}
\affiliation{Moscow Institute of Physics and Technology, Dolgoprudny, Moscow region 141701, Russia}

\author{Alexey~Shilko}
\affiliation{Russian Quantum Center, Skolkovo, Moscow 121205, Russia}
\affiliation{NTI Center for Quantum Communications, National University of Science and Technology MISiS, Moscow 119049, Russia}

\author{Elena~Obraztsova}
\affiliation{Prokhorov General Physics Institute of the Russian Academy of Sciences, Moscow 119991, Russia}
\affiliation{Moscow Institute of Physics and Technology, Dolgoprudny, Moscow region 141701, Russia}

\author{Boris~Galagan}
\affiliation{Prokhorov General Physics Institute of the Russian Academy of Sciences, Moscow 119991, Russia}

\author{Vadim~Makarov}
\affiliation{Russian Quantum Center, Skolkovo, Moscow 121205, Russia}
\affiliation{Vigo Quantum Communication Center, University of Vigo, Vigo E-36310, Spain}
\affiliation{NTI Center for Quantum Communications, National University of Science and Technology MISiS, Moscow 119049, Russia}

\date{\today}

\begin{abstract}
We experimentally demonstrate a power limiter based on single-walled carbon nanotubes dispersed in a polymer matrix. This simple fiber-optic device permanently increases its attenuation when subjected to 50\nobreakdash-mW or higher cw illumination at 1550~nm and initiates a fiber-fuse effect at 1 to 5~W. It may be used for protecting quantum key distribution equipment from light-injection attacks. We demonstrate its compatibility with phase- and polarisation-encoding quantum key distribution systems.
\end{abstract}


\maketitle

\section{Introduction}
\label{sec:intro}

Quantum key distribution (QKD) offers a technology for distributing secure keys that essentially is generating a true-random bit subsequence shared between legitimate users \cite{bennett1984,gisin2002,scarani2009}. It is in principle unhackable through computation, and therefore is in the focus of intensive developments. Nevertheless, its practical implementations have a number of vulnerabilities related to the side channels in real hardware. Although some QKD schemes, such as measurement-device-independent one \cite{lo2012}, eliminate vulnerabilities in the photon receiver, the photon source remains vulnerable.

In particular, it is possible to carry out attacks by injecting external laser emission into QKD transmitter. Several types of attacks are possible, such as Trojan-horse \cite{lucamarini2015}, laser-seeding \cite{huang2019,pang2020,lovic2023}, laser-damage \cite{huang2020,ponosova2022}, optical-pumping \cite{fadeev2025}, and induced-photorefraction attack \cite{ye2023,lu2023,han2023}.

The resilience against high-power illumination seems to be a key practical requirement. To mitigate the light-injection attacks, several devices and methods have been proposed, including fiber-optic isolators~\cite{ponosova2022}, optical power limiters based on thermal defocusing~\cite{zhang2021}, photonic-crystal power limiters based on Fano-like resonance~\cite{ghosh2023}, and online monitoring of injection light from the quantum channel~\cite{bethune2000,vakhitov2001,gisin2006,jain2015}.
Unfortunately, they often fail under exposure to high-power cw laser radiation \cite{bugge2014,makarov2016,ponosova2022}, and many of them have not even been tested under other exposure conditions, such as a pulsed laser \cite{peng2024,ponosova2025}.

Here we propose a passive optical ``fuse'' against attacks that inject intense laser radiation. We show that this optical element that is able to provide sufficient protection of equipment from this type of attack. Its key element is carboxymethylcellulose film with dispersed single-walled carbon nanotubes (CMC-CNT)~\cite{chernov2007}. Laser illumination of more than $50~\milli\watt$ induces an irreversible increase in device attenuation, suppressing Eve's power to a level safe for other components behind it. Under high-power laser radiation of $1~\watt$ or higher, it behaves as a one-time fuse that disconnects the communication line. Thus, an attacker will not be able to receive any information about secure keys or damage the internal components of QKD source.

\section{Design of power limiter device}
\label{sec:design}

\begin{figure}
 \includegraphics{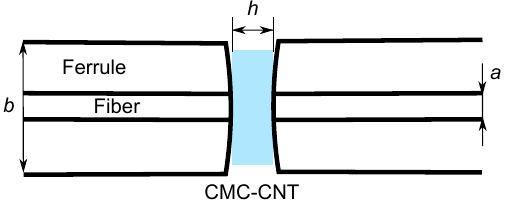}
 \caption{The design of optical fuse (not to scale). The sample of CMC-CNT is between the ferrules of the connectors inserted into a standard bulkhead adapter. Diameter of the fiber $a =125~\micro\meter$, ferrule diameter $b = 2.5~\milli\meter$, sample thickness $h = 5~\micro\meter$.}
 \label{fig:CMC-CNT}
\end{figure}

Our fuse has a simple design that facilitates reproducibility in manual production. It consists of a CMC-CNT composite film that is placed between UPC/FC fiber connectors with single-mode fiber, as shown in~\cref{fig:CMC-CNT}. For experiments, we assembled about seventeen identical samples of this device. 

\begin{figure}
 \includegraphics{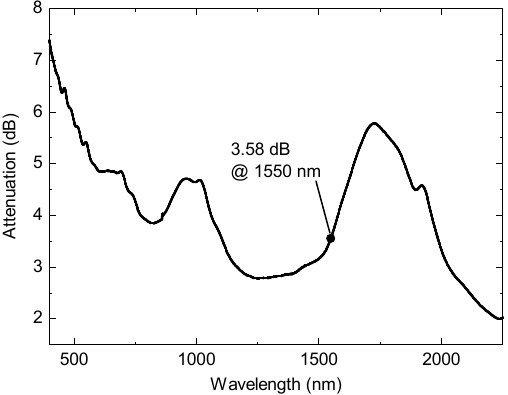}
 \caption{Attenuation spectrum of sheet CMC-CNT film at normal incidence, measured with Perkin-Elmer Lambda 950 spectrophotometer.}
  \label{fig:initial_attenuation}
\end{figure}

The composite film is made of carboxymethyl cellulose with single-walled carbon nanotubes dispersed in it. Carbon nanotubes (CNT) are produced by an arc-discharge method. First, the powder of CNT is dispersed in water solution of carboxymethyl cellulose by ultrasonic dispersion and centrifuged to remove the nanotube bundles and impurities. Then the supernatant (remaining suspension above a solid residue) is poured into a Petri dish and dried. The film produced has a thickness of about $5~\micro\meter$. It shows a homogeneity of structure and properties across its surface. \Cref{fig:initial_attenuation} demonstrates the initial spectrum of typical film's attenuation. The attenuation at a typical QKD operating wavelength of $1550~\nano\meter$ is about $3.6~\deci\bel$.

We cut out sections of the film that match the size of the fiber ferrule with scissors and place them on the tip of the fiber connector with tweezers. To secure the sheets, they are clamped using the second fiber connector within an FC fiber adapter (\cref{fig:CMC-CNT}). The initial attenuation at $1550~\nano\meter$ of the assembled optical fuses ranges  from 3.0 to $8.5~\deci\bel$. It tends to be higher comparing to the initial film attenuation due to extra loss on the gap between the fiber connectors.

\section{Experimental setup}
\label{sec:setup}

\begin{figure}
	\includegraphics{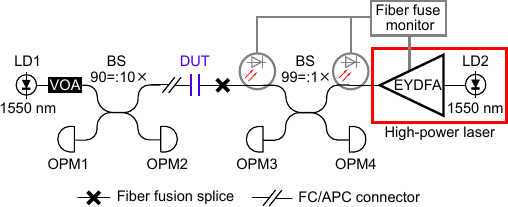}
	\caption{Experimental setup. LD1, laser diode; VOA, variable optical attenuator; OPM, optical power meter; BS, beamsplitter; DUT, device under test; LD2, seed laser diode; EYDFA, erbium-ytterbium-doped fiber amplifier.}
	\label{fig:setup}
\end{figure}

Our experimental setup simulates a hacking scenario in which Eve hacks the QKD source from the quantum channel. \Cref{fig:setup} illustrates the measurement configuration used for testing of optical fuse device. The test sample is exposed to high-power laser (HPL), consisting of a cw $1550~\nano\meter$ seed laser diode LD2 (QPhotonics QFBGLD-1550-100) followed by an erbium-ytterbium-doped fiber amplifier (QGLex custom-made unit \cite{huang2020}). The laser radiation is transmitted through a single-mode fiber to mimic the attack via the quantum channel. As we focus on the effect of optical power on the tested sample, the polarisation of the laser is not characterised. Laser output power can be varied from 0.16 to $5.4~\watt$ at the device under test. Optical power meter 3 (Grandway FHP2B04), monitors the power emitted by the high-power laser in real time. The laser light from HPL transmitted through the sample is continuously monitored by OPM1 (Thorlabs PM400 with S146C sensor). To prevent catastrophic damage to the test equipment by a fiber-fuse effect \cite{kashyap1988, kashyap2013}, a monitor similar to that described in~\cite{huang2020} was used in the setup. Its photodetectors see visible light emitted by the spark traveling through the fiber and turn off EYDFA.

In order to characterise the samples' initial attenuation, a lower power range than that of HPL is required. Before exposing the sample to high power, tests were carried out at lower powers. For this purpose, we use LD1 (Gooch and Housego AA1406) at $50$-$\milli\watt$ power. Its power at the DUT is varied using a manual variable optical attenuator (VOA) placed after LD1 and monitored by OPM2 (Thorlabs PM20CH). The power passing the sample is recorded by OPM4 (Thorlabs PM400 with S154C sensor). 

\begin{table*}
\vspace{-0.7em} 
\caption{Summary of test results.}
\label{tab:samples}
\footnotetext[1]{Small-signal attenuation of assembled device, measured for some samples. All the devices are based on a piece of uniform sheet material with normal-incidence light attenuation of $3.58~\deci\bel$.}
\footnotetext[2]{Samples used with QKD systems, see \cref{sec:QKD}.}
\begin{tblr}[t]{colspec={@{\hskip 1pt}X[11.5mm,r]X[16.5mm,r]X[33mm]X[16.5mm,r]X[14.5mm,r]X[18mm,r]@{\hskip 1pt}},rowspec={X[c,m]|}}
\hline\hline
Sample number & Initial attenuation ($\deci\bel$)\footnotemark[1] & Test exposures ($\milli\watt$); ranges indicate gradual increase of power & Doubling attenuation at ($\milli\watt$) & Fiber fuse occurred at ($\milli\watt$) & Maximum transmitted power ($\milli\watt$) \\
1 & & $145$--$1010$ & $202$ & $1010$ & $6.4$~~~ \\
2 & & $145$--$2030$ & $202$ & $2030$ & $22\,$~~~~~ \\
3 & & $145$--$1270$ & $229$ & $1270$ & $45\,$~~~~~ \\
4 & & $0.5$--$87$, then $145$--$1925$ & $60$ & $1925$ & $4\,$~~~~~ \\
5 & & $0.5$--$87$, then $145$--$5015$ & $60$ & $5015$ & $69\,$~~~~~ \\
6 & & $0.5$--$87$, then $145$--$1010$ & $60$ & $1010$ & $36\,$~~~~~ \\
7 & $3.3$ & $0.01$--$2.6$ & & & \\
8 & $4.8$ & $0.01$--$2.2$ & & & \\
9 & $8.3$ & $0.01$--$2.1$ & & & \\
10 & & $5400$ & & $5400$ & \\
11 & & $5400$ & & $5400$ & $1922\,$~~~~~ \\
12 & & $2720$ & & $2720$ & $1.44\,$ \\
13 & & $1310$, then $5400$ & & $5400$ & $0.21\,$ \\
14 & & $50$ & & & \\
15 & & $500$ & & & \\
16 & &  $0.03\footnotemark[2]$  & & & \\
17 & &  $\footnotemark[2]$  & & & \\
\hline\hline
\end{tblr}
\end{table*}

For most samples, the test procedure is the following. First, the DUT attenuation is characterised using LD1 emission. Next, it is exposed to HPL emission with constant power level starting from its minimum value for at least $5~\minute$, and attenuation of the high-power emission is recorded continuously. If no changes in attenuation are observed, the irradiation power is increased by $1~\deci\bel$. The test is finished if irreversible damage to the fiber occurs or if  power transmitted through the sample falls below OPM1 sensitivity, which is $10~\micro\watt$.  

In addition, we prepared samples exposed only to a single power level for $5~\minute$, to investigate phase and structure changes of material depending on the laser power. Each sample is then examined using optical microscopy and Raman spectroscopy. Raman measurements were performed using LabRam Horiba spectrometer equipped with microscope in back-scattering geometry. The excitation wavelength was $532~\nano\meter$ and spectral resolution was $1~\centi\meter^{-1}$. The microscopic images were recorded using 50-fold and 10-fold objectives.

\section{Test results}
\label{sec:results}

\begin{figure}
	\includegraphics{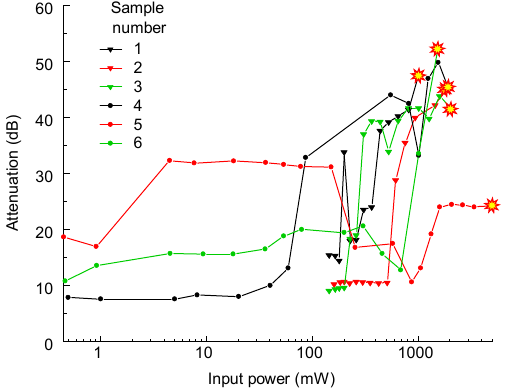}
	\caption{Attenuation of samples 1--6 after their irradiation by increasing amount of power. A spark at the end of the curve denotes the sample initiating the fiber-fuse effect, destroying its fiber pigtail and ending the test.} 
	\label{fig:Att_from_P0}
\end{figure}

\Cref{tab:samples} presents a summary of the tested characteristics of all our samples. The first six samples of the optical fuse were tested under the exposure to laser radiation in the power range from $0.5~\milli\watt$ to more than $1~\watt$. \Cref{fig:Att_from_P0} shows how their attenuation changed. In all samples, an increase in attenuation was observed starting from a certain threshold power. In the end, all samples initiated the fiber-fuse effect \cite{kashyap1988, kashyap2013}. The appearance of the fiber-fuse effect breaks the quantum channel and thus prevents the possibility of any third-party influence on the radiation source of the QKD system. 

The experiments with samples 1--3 were carried out only under HPL, starting from the power of  $145~\milli\watt$. Exposure time was $30~\minute$ at each power level. In these samples, a rapid irreversible change of attenuation was observed with a real-time recording from OPM1 even at the minimum applied power level of $145~\milli\watt$. Therefore, samples 4--6 were tested starting with a much lower power of $0.5~\milli\watt$. The threshold power at which the attenuation begins to increase varies from $1$ to $40~\milli\watt$ (\cref{fig:Att_from_P0}). Up to about $40~\milli\watt$ the change is reversible (i.e.,\ the sample recovers to its initial attenuation after exposure), while at a higher power the change is irreversible. The changes in attenuation show a large sample-to-sample variation.

\begin{figure}
	\includegraphics{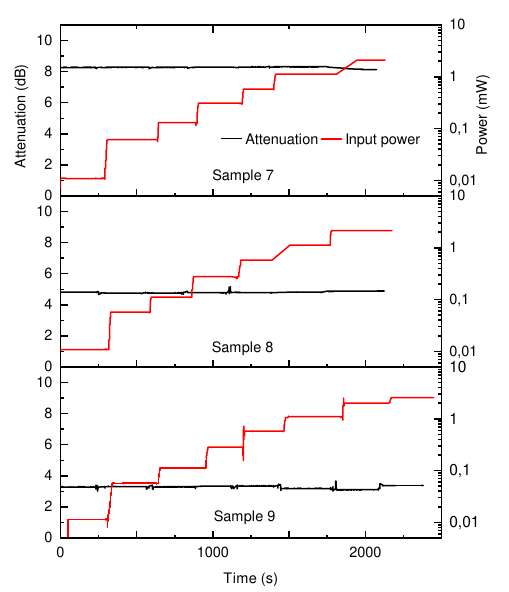}
	\caption{Attenuation of samples exposed to power up to $2.6~\milli\watt$.}
	\label{fig:Low_power}
\end{figure}

To ensure that the optical fuse device will have stable parameters when installed in an operating QKD system, samples 7--9 were tested at low powers in the range from $0.01~\milli\watt$ to $2.6~\milli\watt$. \Cref{fig:Low_power} shows that their attenuation remains constant within the measurement accuracy. Thus, at low powers, the level of attenuation remains stable, while at higher powers the attenuation of the sample increases.

The second important effect for QKD source protection is the initiating of fiber-fuse effect (or travelling optical spark) by the proposed optical fuses. As shown in~\cref{fig:Att_from_P0}, five of six tested samples initiated the fiber fuse at power ranging from 1 to $2~\watt$. These experimentally observed values are in good agreement with theoretical estimations of threshold power of fiber-fuse effect initiation in the case of contamination by carbon-containing materials~\cite{shuto2016}. According to the model in~\cite{shuto2016}, threshold power is $1.5~\watt$ for carbon-induced attenuation higher than $4~\deci\bel$. The threshold power of fiber-fuse effect increases with decrease in attenuation. It is proven experimentally that there is no fiber-fuse effect even at a power of $3~\watt$ in the case of perfectly clean connectors~\cite{rosa2002}. Therefore, the presence of CMC-CNT at the junction of two fiber-optic connectors reduces the threshold of the fiber fuse initiation compared to clean connectors.

Here, we also note, we also conducted preliminary experiments with dispersed graphite in carboxymethyl cellulose and with pure films of carboxymethyl cellulose without any contamination. However, the fiber-fuse effect was not observed even at an input power of $5.4~\watt$.

The fiber-fuse effect interrupts key generation, protecting confidential information and ensuring safety of QKD. At the same time, it damages the telecommunications line permanently. To avoid the unwanted breakdown of a large section of fiber,  an adiabatic taper can be used after our fuse device~\cite{dianov2004}. This additional element stops spark propagation, thereby keeping the main telecommunication line intact. In our experiments, in order to avoid unwanted damage to the equipment by the spark, the signal from the fiber-fuse monitor turned off HPL.

\begin{figure}
	\includegraphics[width=1\linewidth]{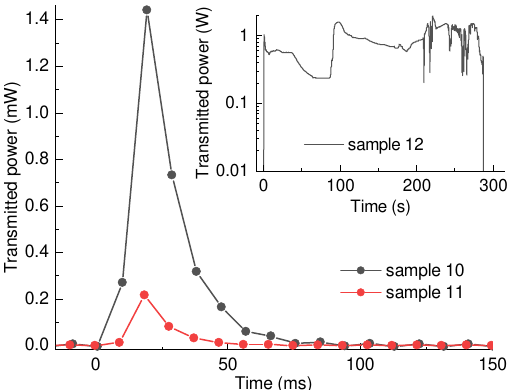}
	\caption{Analysis of the optical fuse response to high-power laser exposure of $5.4~\watt$ for samples 10--12.}
	\label{fig:5.4W}
\end{figure}

Finally, we test optical fuse under extreme conditions by immediately applying the highest available power. Four pristine samples 10--13 were tested under exposure to $5.4~\watt$ laser radiation. \Cref{fig:5.4W} shows the change in the power transmitted through the samples over time. In three samples, we observed an almost instantaneous occurrence of the fiber-fuse effect. Sample~13, which initiated the spark instantly, is not shown in the plots because we forgot to start recording on OPM1. The maximum transmitted power is found to be on the order of a milliwatt. However, the fourth sample transmitted up to $1.9~\watt$ and initiated the spark almost $5~\minute$ into the exposure (see inset in \cref{fig:5.4W}). 

\begin{figure}
	\includegraphics{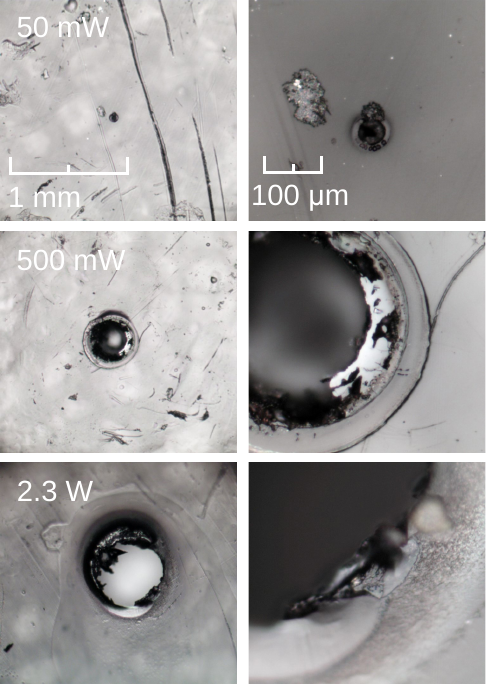}
	\caption{Microscope images of samples 14, 15, and 2. Each row shows one sample after exposure to the denoted power, in two magnifications. Right-hand-side images show the magnified damage spot.}
	\label{fig:film}
\end{figure}

Samples 2, 14, and 15 were used to investigate the microscopic changes in films induced by exposure to different power levels. The microscopic images of the CMC-CNT film after HPL irradiation are shown in \cref{fig:film}. These are photos of film taken after disassembling three samples that were exposed to different power levels. Complete destruction of the structure is observed in the middle of the sample. The film is melted and evaporated under the beam. 

\begin{figure}
	\includegraphics[width=60mm]{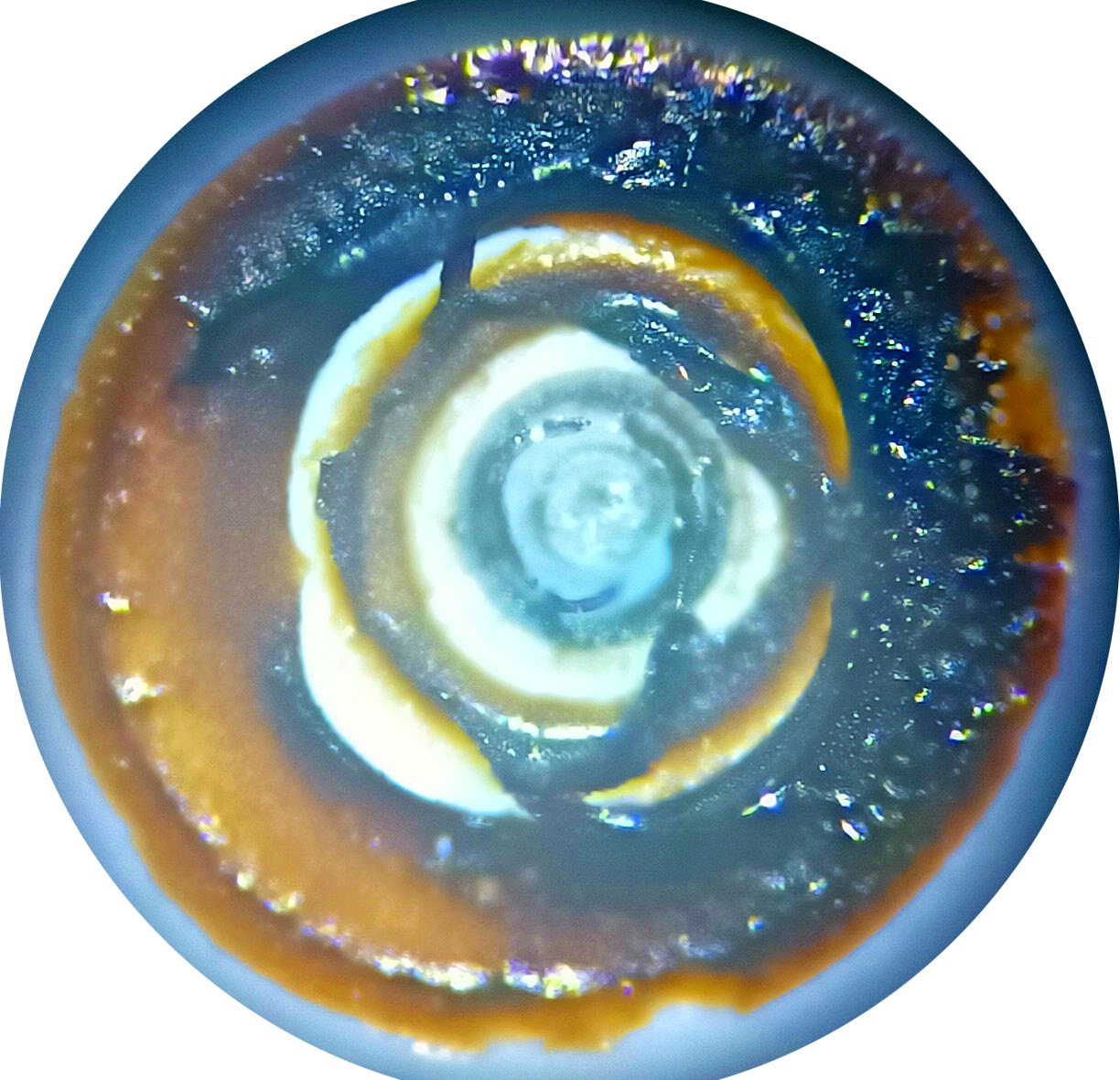}
	\caption{Microscope photo (under oblique illumination) of the $2.5$-$\milli\meter$ diameter ferrule on the ``downwind'' side of sample~1.}
	\label{fig:ferrule}
\end{figure}

The photo in \cref{fig:ferrule} shows how the end of the ferule from sample 1 has changed after exposure to a powerful laser and the occurrence of the fiber-fuse effect. The degradation of the ceramic structure is clearly visible.

\begin{figure}
	\includegraphics{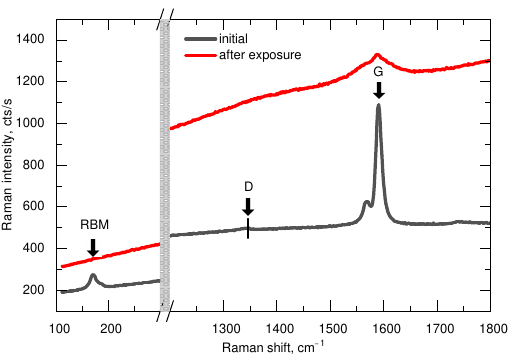}
	\caption{Raman spectra of the initial CMC-CNT film (black) and the irradiated film on the edge of the damaged area in sample~15 after $0.5$-$\watt$ exposure (red). The Raman excitation wavelength was $532~\nano\meter$.}
	\label{fig:raman}
\end{figure}

The Raman spectra of CMC-CNT film before and after the laser damage are presented in~\cref{fig:raman}. If there is no visible damage of the film, the Raman spectra don’t change. They show the vibrational modes typical for single-walled carbon nanotubes: the radial breathing mode (RBM) at $170~\centi\meter^{-1}$, split G-mode at $1591~\centi\meter^{-1}$ and weak D-mode at $1345~\centi\meter^{-1}$.  On the contrary, on the edge of the damaged area all ``fingerprint'' Raman modes of nanotubes disappear. D-mode becomes strongly broadened while G-mode is weakened, broadened, and shifted to smaller wavenumbers. In this case the graphitization of the polymer film takes place, probably together with the degradation and amorphization of CNTs. In addition, the scattering background after the high-power irradiation was slightly higher since the surface of CMC-CNT ceased to be smooth. It indicates that under exposure to HPL, phase transformation of carbon nanotubes into amorphous carbon takes place. 

The base itself, carboxymethyl cellulose, is an optically transparent film in visible and IR range. Heating of the sample occurs only because the nanotubes absorb irradiation. Therefore, it is useless to use the polymer in its pure form. 

\section{Demonstration with quantum key distribution systems}
\label{sec:QKD}

The insertion of material such as CMC-CNT should not affect the operation of communication systems with either polarisation or phase-shift encoding. This hypothesis has been tested in practice.

\begin{figure}
	\includegraphics{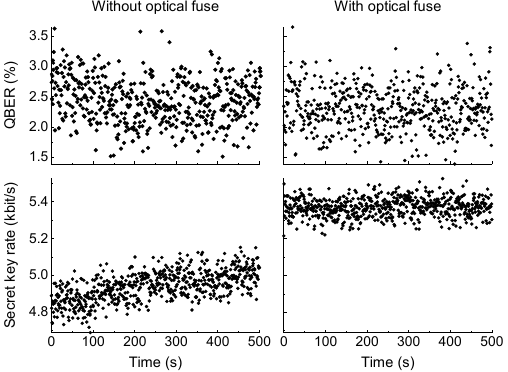}
	\caption{Performance of phase-encoded plug-and-play QKD system with and without our optical fuse device. Sample 16 was used. The quantum channel length was $50~\meter$ of optical fiber.}
	\label{fig:QKD}
\end{figure}

We have tested and compared the operation of two QKD systems with and without optical fuse installed after their transmitters. One system is phase-encoded with a plug-and-play two-pass autocompensating scheme \cite{rodimin2019} and another is polarisation-encoded with prepare-and-measure scheme under development at QRate \cite{makarov2024}. The former runs a plain BB84 protocol and the latter decoy-state BB84 protocol.

In the first system, Alice receives bright optical pulses from Bob, attenuates them to a single photon level, and reflects back. The average optical power incoming from the quantum channel to Alice, and thus impinging on our optical fuse, is about $30~\micro\watt$. Further details of the system operation are recapped in Appendix~\ref{sec:QKD-system-plug-and-play}. To provide similar conditions of system operation in both cases---with and without optical fuse---we adjust the attenuation of Alice's VOA to achieve the same total attenuation in her apparatus. The system software automatically operates the optical scheme and post-processes data to determine the sifted key, its length and quantum bit error ratio (QBER). The experimental results are shown in \cref{fig:QKD}.

\begin{figure}
	\includegraphics{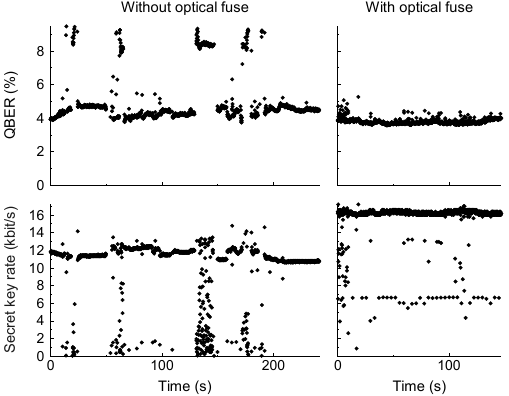}
	\caption{Performance of polarisation-encoded commercial prototype QKD system with and without our optical fuse device. Sample 17 was used. The quantum channel length was a few meters of optical fiber.}
	\label{fig:QKD_QRate}
\end{figure}

The polarisation-encoded prototype system by QRate is described in detail in \cite{makarov2024}. It includes a wavelength-multiplexed optical synchronisation channel at $1554.9~\nano\meter$ applying $10~\nano\watt$ average power to our optical fuse device. Results of a similar test on it are shown in \cref{fig:QKD_QRate}. Fluctuations of the logged quantum bit error ratio (QBER) and key rate are owing to the system's performing internal re-calibrations every minute, which is a normal behaviour of this prototype.

In both tests, the main performance indicators of both systems remain largely unaffected. This confirms that our optical fuse device does not affect their normal operation.

\section{Conclusion}
\label{sec:conclusion}

In this work, we propose a simple design of optical fuse for protecting QKD transmitters against the high-power light-injection attacks. The operation of the device is based on laser-induced changes within CMC-CNT film.

We perform a detailed analysis of laser damage to this fuse device in a wide optical power range from $10~\micro\watt$ to $5.4~\watt$. Our results indicate that the proposed device effectively protects the transmitter against $1550$-$\nano\meter$ cw laser damage through two primary mechanisms:
\begin{itemize}
	\item fiber-fuse effect: by incorporating CMC-CNT film at the junction of two fiber-optic connectors, the threshold for the occurrence of the optical spark is significantly reduced compared to clean connectors;
	\item rise of attenuation: exposure to cw power greater than $50~\milli\watt$ induces an irreversible increase in attenuation of the device.
\end{itemize}

The current limitation of our work is that the attenuation increase varies non-monotonically and has a large sample-to-sample variation. Improving repeatability of this design can be future study.

We have installed our fuse devices into two QKD systems, which has not adversely affected their performance. Therefore, our device might be a good candidate for protection of QKD transmitters against the light-injection attacks.

\begin{figure*}
	\includegraphics[width=0.95\linewidth]{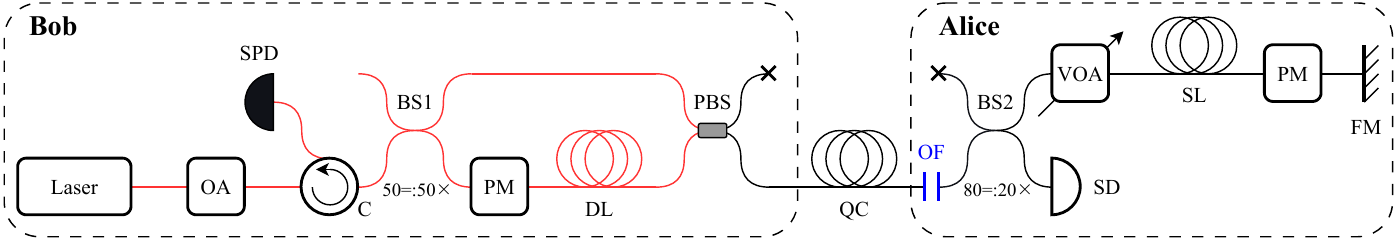}
	\caption{Optical scheme of the plug-and-play QKD system. Black lines are single-mode fiber, red (gray) lines are polarisation-maintaining fiber. OA, optical attenuator; C, circulator; SPD, single-photon detector; BS, beam splitter; PM, phase modulator; DL, delay line; PBS, polarisation beam splitter; QC, quantum channel; OF, optical fuse being tested; VOA, variable optical attenuator; SD, synchronisation detector; SL, storage line; FM, Faraday mirror. Figure is adapted from~\cite{rodimin2019}.}
	\label{QKD_system}
\end{figure*}

\acknowledgments

We thank Vasiliy Koltashev for discussions and advice.
We thank QRate for the opportunity to test our samples on the educational complex and commercial prototype of the QKD system.

{\em Funding:}
Ministry of Education and Science of Russia (program NTI center for quantum communications) and the Russian Science Foundation (grants 21-42-00040 and 24-42-04003). V.M.\ acknowledges funding from MICIN with funding from the European Union NextGenerationEU (PRTR-C17.I1) and the Galician Regional Government with own funding through ``Planes Complementarios de I+D+I con las Comunidades Aut{\' o}nomas'' in Quantum Communication.

{\em Author contributions:}
E.B.\ and A.P.\ performed the experiment and analysed the data. E.B.,\ B.G.,\ and A.P.\ assisted in planning the experiment. N.A.\ prepared and characterised the samples and analysed the data. A.S.\ performed the tests with the QKD systems. E.B.,\ A.P.,\ N.A.,\ and A.S.\ wrote the article with input from all authors. B.G.,\ E.O.,\ and V.M.\ supervised the project. 

\appendix

\section{Operation of plug-and-play QKD system}
\label{sec:QKD-system-plug-and-play}	


\Cref{QKD_system} shows the optical scheme of this system. All the components of Bob are polarisation-maintaining~\cite{rodimin2019}. Bob's laser sends out a burst of pulses (a train), each of which is divided in half by intensity on the beam splitter (BS1). A delay line (DL) is installed in one of the arms of the interferometer, which separates each pair of pulses in time, after which they enter the quantum channel (QC).

At Alice's side, part of the incoming light is diverted to a synchronization detector (SD), which triggers Alice's phase modulator (PM) after a suitable delay. A storage line (SL) is long enough to store the entire train. This greatly reduces noise clicks of Bob's detectors that would otherwise arise from backreflections and Rayleigh backscattering of bright light in the quantum channel coinciding in time with the signal from Alice. A randomly chosen phase modulation of $0$, $\pi/2$, $\pi$, or $3\pi/2$ is applied by Alice to one pulse from each pair both before and after its reflection from a Faraday mirror (FM), creating a phase difference between the pulses. The polarisation of the pulses changes to orthogonal at the reflection by FM, allowing an accurate phase modulation by the polarisation-dependent PM regardless of the pulse polarisation. After being attenuated by two passes through a variable optical attenuator (VOA) to a single-photon level, the pulses enter the quantum channel and return to Bob.

At Bob's polarisation beam splitter (PBS), each pulse is routed to the arm of the interferometer opposite the one it came from. This scheme is implemented with one single-photon detector (SPD). To maintain the security of the BB84 protocol, Bob selects not only the basis but also the detected bit by applying phase shift of $0$, $\pi/2$, $\pi$ or $3\pi/2$ to his PM (i.e.,\ he implements a four-state Bob). Depending on the selected state and basis on Alice's side and the basis on Bob's side, constructive or destructive interference happens at BS1. Because of this, Bob's SPD clicks or does not click.

\bibliography{library}

\begin{thebibliography}{32}%
\makeatletter
\providecommand \@ifxundefined [1]{%
 \@ifx{#1\undefined}
}%
\providecommand \@ifnum [1]{%
 \ifnum #1\expandafter \@firstoftwo
 \else \expandafter \@secondoftwo
 \fi
}%
\providecommand \@ifx [1]{%
 \ifx #1\expandafter \@firstoftwo
 \else \expandafter \@secondoftwo
 \fi
}%
\providecommand \natexlab [1]{#1}%
\providecommand \enquote  [1]{``#1''}%
\providecommand \bibnamefont  [1]{#1}%
\providecommand \bibfnamefont [1]{#1}%
\providecommand \citenamefont [1]{#1}%
\providecommand \href@noop [0]{\@secondoftwo}%
\providecommand \href [0]{\begingroup \@sanitize@url \@href}%
\providecommand \@href[1]{\@@startlink{#1}\@@href}%
\providecommand \@@href[1]{\endgroup#1\@@endlink}%
\providecommand \@sanitize@url [0]{\catcode `\\12\catcode `\$12\catcode
  `\&12\catcode `\#12\catcode `\^12\catcode `\_12\catcode `\%12\relax}%
\providecommand \@@startlink[1]{}%
\providecommand \@@endlink[0]{}%
\providecommand \url  [0]{\begingroup\@sanitize@url \@url }%
\providecommand \@url [1]{\endgroup\@href {#1}{\urlprefix }}%
\providecommand \urlprefix  [0]{URL }%
\providecommand \Eprint [0]{\href }%
\providecommand \doibase [0]{https://doi.org/}%
\providecommand \selectlanguage [0]{\@gobble}%
\providecommand \bibinfo  [0]{\@secondoftwo}%
\providecommand \bibfield  [0]{\@secondoftwo}%
\providecommand \translation [1]{[#1]}%
\providecommand \BibitemOpen [0]{}%
\providecommand \bibitemStop [0]{}%
\providecommand \bibitemNoStop [0]{.\EOS\space}%
\providecommand \EOS [0]{\spacefactor3000\relax}%
\providecommand \BibitemShut  [1]{\csname bibitem#1\endcsname}%
\let\auto@bib@innerbib\@empty
\bibitem [{\citenamefont {Bennett}\ and\ \citenamefont
  {Brassard}(1984)}]{bennett1984}%
  \BibitemOpen
  \bibfield  {author} {\bibinfo {author} {\bibfnamefont {C.~H.}\ \bibnamefont
  {Bennett}}\ and\ \bibinfo {author} {\bibfnamefont {G.}~\bibnamefont
  {Brassard}},\ }\bibfield  {title} {\bibinfo {title} {Quantum cryptography:
  public key distribution and coin tossing},\ }in\ \href@noop {} {\emph
  {\bibinfo {booktitle} {Proc. International Conference on Computers, Systems,
  and Signal Processing}}}\ (\bibinfo  {publisher} {IEEE Press, New York},\
  \bibinfo {address} {Bangalore, India},\ \bibinfo {year} {1984})\ pp.\
  \bibinfo {pages} {175--179}\BibitemShut {NoStop}%
\bibitem [{\citenamefont {Gisin}\ \emph {et~al.}(2002)\citenamefont {Gisin},
  \citenamefont {Ribordy}, \citenamefont {Tittel},\ and\ \citenamefont
  {Zbinden}}]{gisin2002}%
  \BibitemOpen
  \bibfield  {author} {\bibinfo {author} {\bibfnamefont {N.}~\bibnamefont
  {Gisin}}, \bibinfo {author} {\bibfnamefont {G.}~\bibnamefont {Ribordy}},
  \bibinfo {author} {\bibfnamefont {W.}~\bibnamefont {Tittel}},\ and\ \bibinfo
  {author} {\bibfnamefont {H.}~\bibnamefont {Zbinden}},\ }\bibfield  {title}
  {\bibinfo {title} {Quantum cryptography},\ }\href
  {https://doi.org/10.1103/RevModPhys.74.145} {\bibfield  {journal} {\bibinfo
  {journal} {Rev. Mod. Phys.}\ }\textbf {\bibinfo {volume} {74}},\ \bibinfo
  {pages} {145} (\bibinfo {year} {2002})}\BibitemShut {NoStop}%
\bibitem [{\citenamefont {Scarani}\ \emph {et~al.}(2009)\citenamefont
  {Scarani}, \citenamefont {Bechmann-Pasquinucci}, \citenamefont {Cerf},
  \citenamefont {Du\v{s}ek}, \citenamefont {L\"{u}tkenhaus},\ and\
  \citenamefont {Peev}}]{scarani2009}%
  \BibitemOpen
  \bibfield  {author} {\bibinfo {author} {\bibfnamefont {V.}~\bibnamefont
  {Scarani}}, \bibinfo {author} {\bibfnamefont {H.}~\bibnamefont
  {Bechmann-Pasquinucci}}, \bibinfo {author} {\bibfnamefont {N.~J.}\
  \bibnamefont {Cerf}}, \bibinfo {author} {\bibfnamefont {M.}~\bibnamefont
  {Du\v{s}ek}}, \bibinfo {author} {\bibfnamefont {N.}~\bibnamefont
  {L\"{u}tkenhaus}},\ and\ \bibinfo {author} {\bibfnamefont {M.}~\bibnamefont
  {Peev}},\ }\bibfield  {title} {\bibinfo {title} {The security of practical
  quantum key distribution},\ }\href
  {https://doi.org/10.1103/RevModPhys.81.1301} {\bibfield  {journal} {\bibinfo
  {journal} {Rev. Mod. Phys.}\ }\textbf {\bibinfo {volume} {81}},\ \bibinfo
  {eid} {1301} (\bibinfo {year} {2009})}\BibitemShut {NoStop}%
\bibitem [{\citenamefont {Lo}\ \emph {et~al.}(2012)\citenamefont {Lo},
  \citenamefont {Curty},\ and\ \citenamefont {Qi}}]{lo2012}%
  \BibitemOpen
  \bibfield  {author} {\bibinfo {author} {\bibfnamefont {H.-K.}\ \bibnamefont
  {Lo}}, \bibinfo {author} {\bibfnamefont {M.}~\bibnamefont {Curty}},\ and\
  \bibinfo {author} {\bibfnamefont {B.}~\bibnamefont {Qi}},\ }\bibfield
  {title} {\bibinfo {title} {Measurement-device-independent quantum key
  distribution},\ }\href {https://doi.org/10.1103/PhysRevLett.108.130503}
  {\bibfield  {journal} {\bibinfo  {journal} {Phys. Rev. Lett.}\ }\textbf
  {\bibinfo {volume} {108}},\ \bibinfo {pages} {130503} (\bibinfo {year}
  {2012})}\BibitemShut {NoStop}%
\bibitem [{\citenamefont {Lucamarini}\ \emph {et~al.}(2015)\citenamefont
  {Lucamarini}, \citenamefont {Choi}, \citenamefont {Ward}, \citenamefont
  {Dynes}, \citenamefont {Yuan},\ and\ \citenamefont
  {Shields}}]{lucamarini2015}%
  \BibitemOpen
  \bibfield  {author} {\bibinfo {author} {\bibfnamefont {M.}~\bibnamefont
  {Lucamarini}}, \bibinfo {author} {\bibfnamefont {I.}~\bibnamefont {Choi}},
  \bibinfo {author} {\bibfnamefont {M.~B.}\ \bibnamefont {Ward}}, \bibinfo
  {author} {\bibfnamefont {J.~F.}\ \bibnamefont {Dynes}}, \bibinfo {author}
  {\bibfnamefont {Z.~L.}\ \bibnamefont {Yuan}},\ and\ \bibinfo {author}
  {\bibfnamefont {A.~J.}\ \bibnamefont {Shields}},\ }\bibfield  {title}
  {\bibinfo {title} {Practical security bounds against the {T}rojan-horse
  attack in quantum key distribution},\ }\href
  {https://doi.org/10.1103/PhysRevX.5.031030} {\bibfield  {journal} {\bibinfo
  {journal} {Phys. Rev. X}\ }\textbf {\bibinfo {volume} {5}},\ \bibinfo {pages}
  {031030} (\bibinfo {year} {2015})}\BibitemShut {NoStop}%
\bibitem [{\citenamefont {Huang}\ \emph {et~al.}(2019)\citenamefont {Huang},
  \citenamefont {Navarrete}, \citenamefont {Sun}, \citenamefont {Chaiwongkhot},
  \citenamefont {Curty},\ and\ \citenamefont {Makarov}}]{huang2019}%
  \BibitemOpen
  \bibfield  {author} {\bibinfo {author} {\bibfnamefont {A.}~\bibnamefont
  {Huang}}, \bibinfo {author} {\bibfnamefont {{\'A}.}~\bibnamefont
  {Navarrete}}, \bibinfo {author} {\bibfnamefont {S.-H.}\ \bibnamefont {Sun}},
  \bibinfo {author} {\bibfnamefont {P.}~\bibnamefont {Chaiwongkhot}}, \bibinfo
  {author} {\bibfnamefont {M.}~\bibnamefont {Curty}},\ and\ \bibinfo {author}
  {\bibfnamefont {V.}~\bibnamefont {Makarov}},\ }\bibfield  {title} {\bibinfo
  {title} {Laser-seeding attack in quantum key distribution},\ }\href
  {https://doi.org/10.1103/PhysRevApplied.12.064043} {\bibfield  {journal}
  {\bibinfo  {journal} {Phys. Rev. Appl.}\ }\textbf {\bibinfo {volume} {12}},\
  \bibinfo {pages} {064043} (\bibinfo {year} {2019})}\BibitemShut {NoStop}%
\bibitem [{\citenamefont {Pang}\ \emph {et~al.}(2020)\citenamefont {Pang},
  \citenamefont {Yang}, \citenamefont {Zhang}, \citenamefont {Dou},
  \citenamefont {Li}, \citenamefont {Gao},\ and\ \citenamefont
  {Jin}}]{pang2020}%
  \BibitemOpen
  \bibfield  {author} {\bibinfo {author} {\bibfnamefont {X.-L.}\ \bibnamefont
  {Pang}}, \bibinfo {author} {\bibfnamefont {A.-L.}\ \bibnamefont {Yang}},
  \bibinfo {author} {\bibfnamefont {C.-N.}\ \bibnamefont {Zhang}}, \bibinfo
  {author} {\bibfnamefont {J.-P.}\ \bibnamefont {Dou}}, \bibinfo {author}
  {\bibfnamefont {H.}~\bibnamefont {Li}}, \bibinfo {author} {\bibfnamefont
  {J.}~\bibnamefont {Gao}},\ and\ \bibinfo {author} {\bibfnamefont {X.-M.}\
  \bibnamefont {Jin}},\ }\bibfield  {title} {\bibinfo {title} {Hacking quantum
  key distribution via injection locking},\ }\href
  {https://doi.org/10.1103/PhysRevApplied.13.034008} {\bibfield  {journal}
  {\bibinfo  {journal} {Phys. Rev. Appl.}\ }\textbf {\bibinfo {volume} {13}},\
  \bibinfo {pages} {034008} (\bibinfo {year} {2020})}\BibitemShut {NoStop}%
\bibitem [{\citenamefont {Lovic}\ \emph {et~al.}(2023)\citenamefont {Lovic},
  \citenamefont {Marangon}, \citenamefont {Smith}, \citenamefont {Woodward},\
  and\ \citenamefont {Shields}}]{lovic2023}%
  \BibitemOpen
  \bibfield  {author} {\bibinfo {author} {\bibfnamefont {V.}~\bibnamefont
  {Lovic}}, \bibinfo {author} {\bibfnamefont {D.}~\bibnamefont {Marangon}},
  \bibinfo {author} {\bibfnamefont {P.}~\bibnamefont {Smith}}, \bibinfo
  {author} {\bibfnamefont {R.}~\bibnamefont {Woodward}},\ and\ \bibinfo
  {author} {\bibfnamefont {A.}~\bibnamefont {Shields}},\ }\bibfield  {title}
  {\bibinfo {title} {Quantified effects of the laser-seeding attack in quantum
  key distribution},\ }\href {https://doi.org/10.1103/PhysRevApplied.20.044005}
  {\bibfield  {journal} {\bibinfo  {journal} {Phys. Rev. Appl.}\ }\textbf
  {\bibinfo {volume} {20}},\ \bibinfo {pages} {044005} (\bibinfo {year}
  {2023})}\BibitemShut {NoStop}%
\bibitem [{\citenamefont {Huang}\ \emph {et~al.}(2020)\citenamefont {Huang},
  \citenamefont {Li}, \citenamefont {Egorov}, \citenamefont {Tchouragoulov},
  \citenamefont {Kumar},\ and\ \citenamefont {Makarov}}]{huang2020}%
  \BibitemOpen
  \bibfield  {author} {\bibinfo {author} {\bibfnamefont {A.}~\bibnamefont
  {Huang}}, \bibinfo {author} {\bibfnamefont {R.}~\bibnamefont {Li}}, \bibinfo
  {author} {\bibfnamefont {V.}~\bibnamefont {Egorov}}, \bibinfo {author}
  {\bibfnamefont {S.}~\bibnamefont {Tchouragoulov}}, \bibinfo {author}
  {\bibfnamefont {K.}~\bibnamefont {Kumar}},\ and\ \bibinfo {author}
  {\bibfnamefont {V.}~\bibnamefont {Makarov}},\ }\bibfield  {title} {\bibinfo
  {title} {Laser-damage attack against optical attenuators in quantum key
  distribution},\ }\href {https://doi.org/10.1103/PhysRevApplied.13.034017}
  {\bibfield  {journal} {\bibinfo  {journal} {Phys. Rev. Appl.}\ }\textbf
  {\bibinfo {volume} {13}},\ \bibinfo {pages} {034017} (\bibinfo {year}
  {2020})}\BibitemShut {NoStop}%
\bibitem [{\citenamefont {Ponosova}\ \emph {et~al.}(2022)\citenamefont
  {Ponosova}, \citenamefont {Ruzhitskaya}, \citenamefont {Chaiwongkhot},
  \citenamefont {Egorov}, \citenamefont {Makarov},\ and\ \citenamefont
  {Huang}}]{ponosova2022}%
  \BibitemOpen
  \bibfield  {author} {\bibinfo {author} {\bibfnamefont {A.}~\bibnamefont
  {Ponosova}}, \bibinfo {author} {\bibfnamefont {D.}~\bibnamefont
  {Ruzhitskaya}}, \bibinfo {author} {\bibfnamefont {P.}~\bibnamefont
  {Chaiwongkhot}}, \bibinfo {author} {\bibfnamefont {V.}~\bibnamefont
  {Egorov}}, \bibinfo {author} {\bibfnamefont {V.}~\bibnamefont {Makarov}},\
  and\ \bibinfo {author} {\bibfnamefont {A.}~\bibnamefont {Huang}},\ }\bibfield
   {title} {\bibinfo {title} {Protecting fiber-optic quantum key distribution
  sources against light-injection attacks},\ }\href
  {https://doi.org/10.1103/PRXQuantum.3.040307} {\bibfield  {journal} {\bibinfo
   {journal} {PRX Quantum}\ }\textbf {\bibinfo {volume} {3}},\ \bibinfo {pages}
  {040307} (\bibinfo {year} {2022})}\BibitemShut {NoStop}%
\bibitem [{\citenamefont {Fadeev}\ \emph {et~al.}()\citenamefont {Fadeev},
  \citenamefont {Ponosova}, \citenamefont {Peng}, \citenamefont {Huang},
  \citenamefont {Shakhovoy},\ and\ \citenamefont {Makarov}}]{fadeev2025}%
  \BibitemOpen
  \bibfield  {author} {\bibinfo {author} {\bibfnamefont {M.}~\bibnamefont
  {Fadeev}}, \bibinfo {author} {\bibfnamefont {A.}~\bibnamefont {Ponosova}},
  \bibinfo {author} {\bibfnamefont {Q.}~\bibnamefont {Peng}}, \bibinfo {author}
  {\bibfnamefont {A.}~\bibnamefont {Huang}}, \bibinfo {author} {\bibfnamefont
  {R.}~\bibnamefont {Shakhovoy}},\ and\ \bibinfo {author} {\bibfnamefont
  {V.}~\bibnamefont {Makarov}},\ }\bibfield  {title} {\bibinfo {title}
  {Optical-pumping attack on a quantum key distribution laser source},\
  }\href@noop {} {\ }\Eprint {https://arxiv.org/abs/2503.11239}
  {arXiv:2503.11239 [quant-ph]} \BibitemShut {NoStop}%
\bibitem [{\citenamefont {Ye}\ \emph {et~al.}(2023)\citenamefont {Ye},
  \citenamefont {Chen}, \citenamefont {Zhang}, \citenamefont {Lu},
  \citenamefont {Wang}, \citenamefont {Huang}, \citenamefont {Wang},
  \citenamefont {He}, \citenamefont {Yin}, \citenamefont {Guo},\ and\
  \citenamefont {Han}}]{ye2023}%
  \BibitemOpen
  \bibfield  {author} {\bibinfo {author} {\bibfnamefont {P.}~\bibnamefont
  {Ye}}, \bibinfo {author} {\bibfnamefont {W.}~\bibnamefont {Chen}}, \bibinfo
  {author} {\bibfnamefont {G.-W.}\ \bibnamefont {Zhang}}, \bibinfo {author}
  {\bibfnamefont {F.-Y.}\ \bibnamefont {Lu}}, \bibinfo {author} {\bibfnamefont
  {F.-X.}\ \bibnamefont {Wang}}, \bibinfo {author} {\bibfnamefont {G.-Z.}\
  \bibnamefont {Huang}}, \bibinfo {author} {\bibfnamefont {S.}~\bibnamefont
  {Wang}}, \bibinfo {author} {\bibfnamefont {D.-Y.}\ \bibnamefont {He}},
  \bibinfo {author} {\bibfnamefont {Z.-Q.}\ \bibnamefont {Yin}}, \bibinfo
  {author} {\bibfnamefont {G.-C.}\ \bibnamefont {Guo}},\ and\ \bibinfo {author}
  {\bibfnamefont {Z.-F.}\ \bibnamefont {Han}},\ }\bibfield  {title} {\bibinfo
  {title} {Induced-photorefraction attack against quantum key distribution},\
  }\href {https://doi.org/10.1103/PhysRevApplied.19.054052} {\bibfield
  {journal} {\bibinfo  {journal} {Phys. Rev. Appl.}\ }\textbf {\bibinfo
  {volume} {19}},\ \bibinfo {pages} {054052} (\bibinfo {year}
  {2023})}\BibitemShut {NoStop}%
\bibitem [{\citenamefont {Lu}\ \emph {et~al.}(2023)\citenamefont {Lu},
  \citenamefont {Ye}, \citenamefont {Wang}, \citenamefont {Wang}, \citenamefont
  {Yin}, \citenamefont {Wang}, \citenamefont {Huang}, \citenamefont {Chen},
  \citenamefont {He}, \citenamefont {Fan-Yuan}, \citenamefont {Guo},\ and\
  \citenamefont {Han}}]{lu2023}%
  \BibitemOpen
  \bibfield  {author} {\bibinfo {author} {\bibfnamefont {F.-Y.}\ \bibnamefont
  {Lu}}, \bibinfo {author} {\bibfnamefont {P.}~\bibnamefont {Ye}}, \bibinfo
  {author} {\bibfnamefont {Z.-H.}\ \bibnamefont {Wang}}, \bibinfo {author}
  {\bibfnamefont {S.}~\bibnamefont {Wang}}, \bibinfo {author} {\bibfnamefont
  {Z.-Q.}\ \bibnamefont {Yin}}, \bibinfo {author} {\bibfnamefont
  {R.}~\bibnamefont {Wang}}, \bibinfo {author} {\bibfnamefont {X.-J.}\
  \bibnamefont {Huang}}, \bibinfo {author} {\bibfnamefont {W.}~\bibnamefont
  {Chen}}, \bibinfo {author} {\bibfnamefont {D.-Y.}\ \bibnamefont {He}},
  \bibinfo {author} {\bibfnamefont {G.-J.}\ \bibnamefont {Fan-Yuan}}, \bibinfo
  {author} {\bibfnamefont {G.-C.}\ \bibnamefont {Guo}},\ and\ \bibinfo {author}
  {\bibfnamefont {Z.-F.}\ \bibnamefont {Han}},\ }\bibfield  {title} {\bibinfo
  {title} {Hacking measurement-device-independent quantum key distribution},\
  }\href {https://doi.org/10.1364/OPTICA.485389} {\bibfield  {journal}
  {\bibinfo  {journal} {Optica}\ }\textbf {\bibinfo {volume} {10}},\ \bibinfo
  {pages} {520} (\bibinfo {year} {2023})}\BibitemShut {NoStop}%
\bibitem [{\citenamefont {Han}\ \emph {et~al.}(2023)\citenamefont {Han},
  \citenamefont {Li}, \citenamefont {Tan}, \citenamefont {Zhang}, \citenamefont
  {Cai}, \citenamefont {Yin}, \citenamefont {Ren}, \citenamefont {Xu},
  \citenamefont {Liao},\ and\ \citenamefont {Peng}}]{han2023}%
  \BibitemOpen
  \bibfield  {author} {\bibinfo {author} {\bibfnamefont {L.}~\bibnamefont
  {Han}}, \bibinfo {author} {\bibfnamefont {Y.}~\bibnamefont {Li}}, \bibinfo
  {author} {\bibfnamefont {H.}~\bibnamefont {Tan}}, \bibinfo {author}
  {\bibfnamefont {W.}~\bibnamefont {Zhang}}, \bibinfo {author} {\bibfnamefont
  {W.}~\bibnamefont {Cai}}, \bibinfo {author} {\bibfnamefont {J.}~\bibnamefont
  {Yin}}, \bibinfo {author} {\bibfnamefont {J.}~\bibnamefont {Ren}}, \bibinfo
  {author} {\bibfnamefont {F.}~\bibnamefont {Xu}}, \bibinfo {author}
  {\bibfnamefont {S.}~\bibnamefont {Liao}},\ and\ \bibinfo {author}
  {\bibfnamefont {C.}~\bibnamefont {Peng}},\ }\bibfield  {title} {\bibinfo
  {title} {Effect of light injection on the security of practical quantum key
  distribution},\ }\href {https://doi.org/10.1103/PhysRevApplied.20.044013}
  {\bibfield  {journal} {\bibinfo  {journal} {Phys. Rev. Appl.}\ }\textbf
  {\bibinfo {volume} {20}},\ \bibinfo {pages} {044013} (\bibinfo {year}
  {2023})}\BibitemShut {NoStop}%
\bibitem [{\citenamefont {Zhang}\ \emph {et~al.}(2021)\citenamefont {Zhang},
  \citenamefont {Primaatmaja}, \citenamefont {Haw}, \citenamefont {Gong},
  \citenamefont {Wang},\ and\ \citenamefont {Lim}}]{zhang2021}%
  \BibitemOpen
  \bibfield  {author} {\bibinfo {author} {\bibfnamefont {G.}~\bibnamefont
  {Zhang}}, \bibinfo {author} {\bibfnamefont {I.~W.}\ \bibnamefont
  {Primaatmaja}}, \bibinfo {author} {\bibfnamefont {J.~Y.}\ \bibnamefont
  {Haw}}, \bibinfo {author} {\bibfnamefont {X.}~\bibnamefont {Gong}}, \bibinfo
  {author} {\bibfnamefont {C.}~\bibnamefont {Wang}},\ and\ \bibinfo {author}
  {\bibfnamefont {C.~C.~W.}\ \bibnamefont {Lim}},\ }\bibfield  {title}
  {\bibinfo {title} {Securing practical quantum communication systems with
  optical power limiters},\ }\href
  {https://doi.org/10.1103/PRXQuantum.2.030304} {\bibfield  {journal} {\bibinfo
   {journal} {PRX Quantum}\ }\textbf {\bibinfo {volume} {2}},\ \bibinfo {pages}
  {030304} (\bibinfo {year} {2021})}\BibitemShut {NoStop}%
\bibitem [{\citenamefont {Ghosh}\ \emph {et~al.}(2023)\citenamefont {Ghosh},
  \citenamefont {Das}, \citenamefont {Sharma},\ and\ \citenamefont
  {Medhekar}}]{ghosh2023}%
  \BibitemOpen
  \bibfield  {author} {\bibinfo {author} {\bibfnamefont {N.}~\bibnamefont
  {Ghosh}}, \bibinfo {author} {\bibfnamefont {A.}~\bibnamefont {Das}}, \bibinfo
  {author} {\bibfnamefont {P.}~\bibnamefont {Sharma}},\ and\ \bibinfo {author}
  {\bibfnamefont {S.}~\bibnamefont {Medhekar}},\ }\bibfield  {title} {\bibinfo
  {title} {Photonic crystal power limiter based on fano-like resonance},\
  }\href {https://doi.org/10.1007/s12596-023-01460-y} {\bibfield  {journal}
  {\bibinfo  {journal} {J. Opt.}\ }\textbf {\bibinfo {volume} {53}},\ \bibinfo
  {pages} {2091–2097} (\bibinfo {year} {2023})}\BibitemShut {NoStop}%
\bibitem [{\citenamefont {Bethune}\ and\ \citenamefont
  {Risk}(2000)}]{bethune2000}%
  \BibitemOpen
  \bibfield  {author} {\bibinfo {author} {\bibfnamefont {D.~S.}\ \bibnamefont
  {Bethune}}\ and\ \bibinfo {author} {\bibfnamefont {W.~P.}\ \bibnamefont
  {Risk}},\ }\bibfield  {title} {\bibinfo {title} {An autocompensating
  fiber-optic quantum cryptography system based on polarization splitting of
  light},\ }\href {https://doi.org/10.1109/3.825881} {\bibfield  {journal}
  {\bibinfo  {journal} {IEEE J. Quantum Electron.}\ }\textbf {\bibinfo {volume}
  {36}},\ \bibinfo {pages} {340} (\bibinfo {year} {2000})}\BibitemShut
  {NoStop}%
\bibitem [{\citenamefont {Vakhitov}\ \emph {et~al.}(2001)\citenamefont
  {Vakhitov}, \citenamefont {Makarov},\ and\ \citenamefont
  {Hjelme}}]{vakhitov2001}%
  \BibitemOpen
  \bibfield  {author} {\bibinfo {author} {\bibfnamefont {A.}~\bibnamefont
  {Vakhitov}}, \bibinfo {author} {\bibfnamefont {V.}~\bibnamefont {Makarov}},\
  and\ \bibinfo {author} {\bibfnamefont {D.~R.}\ \bibnamefont {Hjelme}},\
  }\bibfield  {title} {\bibinfo {title} {Large pulse attack as a method of
  conventional optical eavesdropping in quantum cryptography},\ }\href
  {https://doi.org/10.1080/09500340108240904} {\bibfield  {journal} {\bibinfo
  {journal} {J. Mod. Opt.}\ }\textbf {\bibinfo {volume} {48}},\ \bibinfo
  {pages} {2023} (\bibinfo {year} {2001})}\BibitemShut {NoStop}%
\bibitem [{\citenamefont {Gisin}\ \emph {et~al.}(2006)\citenamefont {Gisin},
  \citenamefont {Fasel}, \citenamefont {Kraus}, \citenamefont {Zbinden},\ and\
  \citenamefont {Ribordy}}]{gisin2006}%
  \BibitemOpen
  \bibfield  {author} {\bibinfo {author} {\bibfnamefont {N.}~\bibnamefont
  {Gisin}}, \bibinfo {author} {\bibfnamefont {S.}~\bibnamefont {Fasel}},
  \bibinfo {author} {\bibfnamefont {B.}~\bibnamefont {Kraus}}, \bibinfo
  {author} {\bibfnamefont {H.}~\bibnamefont {Zbinden}},\ and\ \bibinfo {author}
  {\bibfnamefont {G.}~\bibnamefont {Ribordy}},\ }\bibfield  {title} {\bibinfo
  {title} {Trojan-horse attacks on quantum-key-distribution systems},\ }\href
  {https://doi.org/10.1103/PhysRevA.73.022320} {\bibfield  {journal} {\bibinfo
  {journal} {Phys. Rev. A}\ }\textbf {\bibinfo {volume} {73}},\ \bibinfo
  {pages} {022320} (\bibinfo {year} {2006})}\BibitemShut {NoStop}%
\bibitem [{\citenamefont {Jain}\ \emph {et~al.}(2015)\citenamefont {Jain},
  \citenamefont {Stiller}, \citenamefont {Khan}, \citenamefont {Makarov},
  \citenamefont {Marquardt},\ and\ \citenamefont {Leuch}}]{jain2015}%
  \BibitemOpen
  \bibfield  {author} {\bibinfo {author} {\bibfnamefont {N.}~\bibnamefont
  {Jain}}, \bibinfo {author} {\bibfnamefont {B.}~\bibnamefont {Stiller}},
  \bibinfo {author} {\bibfnamefont {I.}~\bibnamefont {Khan}}, \bibinfo {author}
  {\bibfnamefont {V.}~\bibnamefont {Makarov}}, \bibinfo {author} {\bibfnamefont
  {C.}~\bibnamefont {Marquardt}},\ and\ \bibinfo {author} {\bibfnamefont
  {G.}~\bibnamefont {Leuch}},\ }\bibfield  {title} {\bibinfo {title} {Risk
  analysis of {T}rojan-horse attacks on practical quantum key distribution
  systems},\ }\href {https://doi.org/10.1109/JSTQE.2014.2365585} {\bibfield
  {journal} {\bibinfo  {journal} {IEEE J. Sel. Top. Quantum Electron.}\
  }\textbf {\bibinfo {volume} {21}},\ \bibinfo {pages} {6600710} (\bibinfo
  {year} {2015})}\BibitemShut {NoStop}%
\bibitem [{\citenamefont {Bugge}\ \emph {et~al.}(2014)\citenamefont {Bugge},
  \citenamefont {Sauge}, \citenamefont {Ghazali}, \citenamefont {Skaar},
  \citenamefont {Lydersen},\ and\ \citenamefont {Makarov}}]{bugge2014}%
  \BibitemOpen
  \bibfield  {author} {\bibinfo {author} {\bibfnamefont {A.~N.}\ \bibnamefont
  {Bugge}}, \bibinfo {author} {\bibfnamefont {S.}~\bibnamefont {Sauge}},
  \bibinfo {author} {\bibfnamefont {A.~M.~M.}\ \bibnamefont {Ghazali}},
  \bibinfo {author} {\bibfnamefont {J.}~\bibnamefont {Skaar}}, \bibinfo
  {author} {\bibfnamefont {L.}~\bibnamefont {Lydersen}},\ and\ \bibinfo
  {author} {\bibfnamefont {V.}~\bibnamefont {Makarov}},\ }\bibfield  {title}
  {\bibinfo {title} {Laser damage helps the eavesdropper in quantum
  cryptography},\ }\href {https://doi.org/10.1103/PhysRevLett.112.070503}
  {\bibfield  {journal} {\bibinfo  {journal} {Phys. Rev. Lett.}\ }\textbf
  {\bibinfo {volume} {112}},\ \bibinfo {pages} {070503} (\bibinfo {year}
  {2014})}\BibitemShut {NoStop}%
\bibitem [{\citenamefont {Makarov}\ \emph {et~al.}(2016)\citenamefont
  {Makarov}, \citenamefont {Bourgoin}, \citenamefont {Chaiwongkhot},
  \citenamefont {Gagn{\'e}}, \citenamefont {Jennewein}, \citenamefont {Kaiser},
  \citenamefont {Kashyap}, \citenamefont {Legr{\'e}}, \citenamefont
  {Minshull},\ and\ \citenamefont {Sajeed}}]{makarov2016}%
  \BibitemOpen
  \bibfield  {author} {\bibinfo {author} {\bibfnamefont {V.}~\bibnamefont
  {Makarov}}, \bibinfo {author} {\bibfnamefont {J.-P.}\ \bibnamefont
  {Bourgoin}}, \bibinfo {author} {\bibfnamefont {P.}~\bibnamefont
  {Chaiwongkhot}}, \bibinfo {author} {\bibfnamefont {M.}~\bibnamefont
  {Gagn{\'e}}}, \bibinfo {author} {\bibfnamefont {T.}~\bibnamefont
  {Jennewein}}, \bibinfo {author} {\bibfnamefont {S.}~\bibnamefont {Kaiser}},
  \bibinfo {author} {\bibfnamefont {R.}~\bibnamefont {Kashyap}}, \bibinfo
  {author} {\bibfnamefont {M.}~\bibnamefont {Legr{\'e}}}, \bibinfo {author}
  {\bibfnamefont {C.}~\bibnamefont {Minshull}},\ and\ \bibinfo {author}
  {\bibfnamefont {S.}~\bibnamefont {Sajeed}},\ }\bibfield  {title} {\bibinfo
  {title} {Creation of backdoors in quantum communications via laser damage},\
  }\href {https://doi.org/10.1103/PhysRevA.94.030302} {\bibfield  {journal}
  {\bibinfo  {journal} {Phys. Rev. A}\ }\textbf {\bibinfo {volume} {94}},\
  \bibinfo {pages} {030302} (\bibinfo {year} {2016})}\BibitemShut {NoStop}%
\bibitem [{\citenamefont {Peng}\ \emph {et~al.}(2024)\citenamefont {Peng},
  \citenamefont {Gao}, \citenamefont {Zaitsev}, \citenamefont {Wang},
  \citenamefont {Ding}, \citenamefont {Liu}, \citenamefont {Liao},
  \citenamefont {Guo}, \citenamefont {Huang},\ and\ \citenamefont
  {Wu}}]{peng2024}%
  \BibitemOpen
  \bibfield  {author} {\bibinfo {author} {\bibfnamefont {Q.}~\bibnamefont
  {Peng}}, \bibinfo {author} {\bibfnamefont {B.}~\bibnamefont {Gao}}, \bibinfo
  {author} {\bibfnamefont {K.}~\bibnamefont {Zaitsev}}, \bibinfo {author}
  {\bibfnamefont {D.}~\bibnamefont {Wang}}, \bibinfo {author} {\bibfnamefont
  {J.}~\bibnamefont {Ding}}, \bibinfo {author} {\bibfnamefont {Y.}~\bibnamefont
  {Liu}}, \bibinfo {author} {\bibfnamefont {Q.}~\bibnamefont {Liao}}, \bibinfo
  {author} {\bibfnamefont {Y.}~\bibnamefont {Guo}}, \bibinfo {author}
  {\bibfnamefont {A.}~\bibnamefont {Huang}},\ and\ \bibinfo {author}
  {\bibfnamefont {J.}~\bibnamefont {Wu}},\ }\bibfield  {title} {\bibinfo
  {title} {Security boundaries of an optical-power limiter for protecting
  quantum-key-distribution systems},\ }\href
  {https://doi.org/10.1103/PhysRevApplied.21.014026} {\bibfield  {journal}
  {\bibinfo  {journal} {Phys. Rev. Appl.}\ }\textbf {\bibinfo {volume} {21}},\
  \bibinfo {pages} {014026} (\bibinfo {year} {2024})}\BibitemShut {NoStop}%
\bibitem [{\citenamefont {Chernov}\ \emph {et~al.}(2007)\citenamefont
  {Chernov}, \citenamefont {Obraztsova},\ and\ \citenamefont
  {Lobach}}]{chernov2007}%
  \BibitemOpen
  \bibfield  {author} {\bibinfo {author} {\bibfnamefont {A.~I.}\ \bibnamefont
  {Chernov}}, \bibinfo {author} {\bibfnamefont {E.~D.}\ \bibnamefont
  {Obraztsova}},\ and\ \bibinfo {author} {\bibfnamefont {A.~S.}\ \bibnamefont
  {Lobach}},\ }\bibfield  {title} {\bibinfo {title} {Optical properties of
  polymer films with embedded single-wall carbon nanotubes},\ }\href
  {https://doi.org/10.1002/pssb.200776152} {\bibfield  {journal} {\bibinfo
  {journal} {Phys. Stat. Sol. B}\ }\textbf {\bibinfo {volume} {224}},\ \bibinfo
  {pages} {4231–4235} (\bibinfo {year} {2007})}\BibitemShut {NoStop}%
\bibitem [{\citenamefont {Kashyap}\ and\ \citenamefont
  {Blow}(1988)}]{kashyap1988}%
  \BibitemOpen
  \bibfield  {author} {\bibinfo {author} {\bibfnamefont {R.}~\bibnamefont
  {Kashyap}}\ and\ \bibinfo {author} {\bibfnamefont {K.~J.}\ \bibnamefont
  {Blow}},\ }\bibfield  {title} {\bibinfo {title} {Observation of catastrophic
  self-propelled self-focusing in optical fibres},\ }\href
  {https://doi.org/10.1049/el:19880032} {\bibfield  {journal} {\bibinfo
  {journal} {Electron. Lett.}\ }\textbf {\bibinfo {volume} {24}},\ \bibinfo
  {pages} {47} (\bibinfo {year} {1988})}\BibitemShut {NoStop}%
\bibitem [{\citenamefont {Kashyap}(2013)}]{kashyap2013}%
  \BibitemOpen
  \bibfield  {author} {\bibinfo {author} {\bibfnamefont {R.}~\bibnamefont
  {Kashyap}},\ }\bibfield  {title} {\bibinfo {title} {The fiber fuse - from a
  curious effect to a critical issue: {A} 25\textsuperscript{th} year
  retrospective},\ }\href {https://doi.org/10.1364/OE.21.006422} {\bibfield
  {journal} {\bibinfo  {journal} {Opt. Express}\ }\textbf {\bibinfo {volume}
  {19}},\ \bibinfo {pages} {6422} (\bibinfo {year} {2013})}\BibitemShut
  {NoStop}%
\bibitem [{\citenamefont {Shuto}(2016)}]{shuto2016}%
  \BibitemOpen
  \bibfield  {author} {\bibinfo {author} {\bibfnamefont {Y.}~\bibnamefont
  {Shuto}},\ }\bibfield  {title} {\bibinfo {title} {End face damage and fiber
  fuse phenomena in single-mode fiber-optic connectors},\ }\href
  {https://doi.org/10.1155/2016/2781392} {\bibfield  {journal} {\bibinfo
  {journal} {J. Photonics}\ }\textbf {\bibinfo {volume} {11}},\ \bibinfo
  {pages} {1} (\bibinfo {year} {2016})}\BibitemShut {NoStop}%
\bibitem [{\citenamefont {Rosa}\ \emph {et~al.}(2002)\citenamefont {Rosa},
  \citenamefont {Carberry}, \citenamefont {Bhagavatula}, \citenamefont
  {Wagner},\ and\ \citenamefont {Saravanos}}]{rosa2002}%
  \BibitemOpen
  \bibfield  {author} {\bibinfo {author} {\bibfnamefont {M.~D.}\ \bibnamefont
  {Rosa}}, \bibinfo {author} {\bibfnamefont {J.}~\bibnamefont {Carberry}},
  \bibinfo {author} {\bibfnamefont {V.}~\bibnamefont {Bhagavatula}}, \bibinfo
  {author} {\bibfnamefont {K.}~\bibnamefont {Wagner}},\ and\ \bibinfo {author}
  {\bibfnamefont {C.}~\bibnamefont {Saravanos}},\ }\bibfield  {title} {\bibinfo
  {title} {High-power performance of single-mode fiber-optic connectors},\
  }\href {https://doi.org/10.1109/jlt.2002.1007944} {\bibfield  {journal}
  {\bibinfo  {journal} {Lightwave Technol.}\ }\textbf {\bibinfo {volume}
  {11}},\ \bibinfo {pages} {879} (\bibinfo {year} {2002})}\BibitemShut
  {NoStop}%
\bibitem [{\citenamefont {Dianov}\ \emph {et~al.}()\citenamefont {Dianov},
  \citenamefont {Bufetov},\ and\ \citenamefont {Frolov}}]{dianov2004}%
  \BibitemOpen
  \bibfield  {author} {\bibinfo {author} {\bibfnamefont {E.~M.}\ \bibnamefont
  {Dianov}}, \bibinfo {author} {\bibfnamefont {I.~A.}\ \bibnamefont
  {Bufetov}},\ and\ \bibinfo {author} {\bibfnamefont {A.~A.}\ \bibnamefont
  {Frolov}},\ }\href@noop {} {\bibinfo {title} {Device for protecting
  fiber-optic lines against destruction by laser emission}},\ \bibinfo {note}
  {patent RU 2 229 770 C2 (filed 12 Jul 2002, published 27 May
  2007)}\BibitemShut {NoStop}%
\bibitem [{\citenamefont {Rodimin}\ \emph {et~al.}(2019)\citenamefont
  {Rodimin}, \citenamefont {Kiktenko}, \citenamefont {Usova}, \citenamefont
  {Ponomarev}, \citenamefont {Kazieva}, \citenamefont {Miller}, \citenamefont
  {Sokolov}, \citenamefont {Kanapin}, \citenamefont {Losev}, \citenamefont
  {Trushechkin}, \citenamefont {Anufriev}, \citenamefont {Pozhar},
  \citenamefont {Kurochkin}, \citenamefont {Kurochkin},\ and\ \citenamefont
  {Fedorov}}]{rodimin2019}%
  \BibitemOpen
  \bibfield  {author} {\bibinfo {author} {\bibfnamefont {V.~E.}\ \bibnamefont
  {Rodimin}}, \bibinfo {author} {\bibfnamefont {E.~O.}\ \bibnamefont
  {Kiktenko}}, \bibinfo {author} {\bibfnamefont {V.~V.}\ \bibnamefont {Usova}},
  \bibinfo {author} {\bibfnamefont {M.~Y.}\ \bibnamefont {Ponomarev}}, \bibinfo
  {author} {\bibfnamefont {T.~V.}\ \bibnamefont {Kazieva}}, \bibinfo {author}
  {\bibfnamefont {A.~V.}\ \bibnamefont {Miller}}, \bibinfo {author}
  {\bibfnamefont {A.~S.}\ \bibnamefont {Sokolov}}, \bibinfo {author}
  {\bibfnamefont {A.~A.}\ \bibnamefont {Kanapin}}, \bibinfo {author}
  {\bibfnamefont {A.~V.}\ \bibnamefont {Losev}}, \bibinfo {author}
  {\bibfnamefont {A.~S.}\ \bibnamefont {Trushechkin}}, \bibinfo {author}
  {\bibfnamefont {M.~N.}\ \bibnamefont {Anufriev}}, \bibinfo {author}
  {\bibfnamefont {N.~O.}\ \bibnamefont {Pozhar}}, \bibinfo {author}
  {\bibfnamefont {V.~L.}\ \bibnamefont {Kurochkin}}, \bibinfo {author}
  {\bibfnamefont {Y.~V.}\ \bibnamefont {Kurochkin}},\ and\ \bibinfo {author}
  {\bibfnamefont {A.~K.}\ \bibnamefont {Fedorov}},\ }\bibfield  {title}
  {\bibinfo {title} {Modular quantum key distribution setup for research and
  development applications},\ }\href
  {https://doi.org/10.1007/s10946-019-09793-5} {\bibfield  {journal} {\bibinfo
  {journal} {J. Russ. Laser Res.}\ }\textbf {\bibinfo {volume} {40}},\ \bibinfo
  {pages} {221} (\bibinfo {year} {2019})}\BibitemShut {NoStop}%
\bibitem [{\citenamefont {Makarov}\ \emph {et~al.}(2024)\citenamefont
  {Makarov}, \citenamefont {Abrikosov}, \citenamefont {Chaiwongkhot},
  \citenamefont {Fedorov}, \citenamefont {Huang}, \citenamefont {Kiktenko},
  \citenamefont {Petrov}, \citenamefont {Ponosova}, \citenamefont
  {Ruzhitskaya}, \citenamefont {Tayduganov}, \citenamefont {Trefilov},\ and\
  \citenamefont {Zaitsev}}]{makarov2024}%
  \BibitemOpen
  \bibfield  {author} {\bibinfo {author} {\bibfnamefont {V.}~\bibnamefont
  {Makarov}}, \bibinfo {author} {\bibfnamefont {A.}~\bibnamefont {Abrikosov}},
  \bibinfo {author} {\bibfnamefont {P.}~\bibnamefont {Chaiwongkhot}}, \bibinfo
  {author} {\bibfnamefont {A.~K.}\ \bibnamefont {Fedorov}}, \bibinfo {author}
  {\bibfnamefont {A.}~\bibnamefont {Huang}}, \bibinfo {author} {\bibfnamefont
  {E.}~\bibnamefont {Kiktenko}}, \bibinfo {author} {\bibfnamefont
  {M.}~\bibnamefont {Petrov}}, \bibinfo {author} {\bibfnamefont
  {A.}~\bibnamefont {Ponosova}}, \bibinfo {author} {\bibfnamefont
  {D.}~\bibnamefont {Ruzhitskaya}}, \bibinfo {author} {\bibfnamefont
  {A.}~\bibnamefont {Tayduganov}}, \bibinfo {author} {\bibfnamefont
  {D.}~\bibnamefont {Trefilov}},\ and\ \bibinfo {author} {\bibfnamefont
  {K.}~\bibnamefont {Zaitsev}},\ }\bibfield  {title} {\bibinfo {title}
  {Preparing a commercial quantum key distribution system for certification
  against implementation loopholes},\ }\href
  {https://doi.org/10.1103/PhysRevApplied.22.044076} {\bibfield  {journal}
  {\bibinfo  {journal} {Phys. Rev. Appl.}\ }\textbf {\bibinfo {volume} {22}},\
  \bibinfo {pages} {044076} (\bibinfo {year} {2024})}\BibitemShut {NoStop}%
\bibitem [{\citenamefont {Duplinskiy}(2019)}]{duplinskiy2019}%
  \BibitemOpen
  \bibfield  {author} {\bibinfo {author} {\bibfnamefont {A.}~\bibnamefont
  {Duplinskiy}},\ }\emph {\bibinfo {title} {Quantum key distribution with
  high-rate polarization encoding}},\ \href@noop {} {Ph.D. thesis},\ \bibinfo
  {school} {Moscow Institute of Physics and Technology} (\bibinfo {year}
  {2019})\BibitemShut {NoStop}%
\end{thebibliography}%

\end{document}